# Determination of chirality and density control of Néel-type skyrmions with in-plane magnetic field

Senfu Zhang[1,#], Junwei Zhang[1,#], Yan Wen[1], Eugene M. Chudnovsky [2,*] and Xixiang Zhang[1,*]

[1]*Physical Science and Engineering Division (PSE), King Abdullah University of Science and Technology (KAUST), Thuwal 23955-6900, Saudi Arabia*

[2]*Physics Department, Lehman College and Graduate School, The City University of New York, 250 Bedford Park Boulevard West, Bronx, New York 10468-1589, USA*

Magnetic skyrmions are topologically protected nanoscale spin textures exhibiting fascinating physical behaviors. Recent observations of room temperature Néel-type skyrmions in magnetic multilayer films are an important step towards their use in ultra-low power devices. We have investigated the magnetization reversal in a $[Pt/Co/Ta]_{11}$ multilayer sample under a tilted magnetic field using the in-situ Lorentz tunneling electron microscopy. On decreasing the magnetic field individual skyrmions appear that subsequently evolve into snake-like structures growing in the direction opposite to the in-plane magnetic field. We show that this unusual relation between the velocity vector and the magnetic field is dominated by the chirality of the Néel-type skyrmions. It allows one to extract the sign of the Dzyaloshinskii-Moriya constant. We also demonstrate that high concentration of skyrmions can be achieved on increasing the in-plane component of the field. Our micromagnetic simulations agree with our experimental results.

[#] Equally contributed to this work.

[*] E-mail address: xixiang.zhang@kaust.edu.sa (Xixiang Zhang),

EUGENE.CHUDNOVSKY@lehman.cuny.edu (Eugene M Chudnovsky)

Magnetic skyrmions are nanoscale, relatively stable chiral spin textures that are protected topologically[1-4]. Owning to their topological property, small size, and high mobility, skyrmions can be manipulated by much lower current densities than current densities needed to manipulate magnetic domains[3,5-9]. Skyrmions have been therefore proposed as promising candidates for the next generation, low power spintronic devices, such as non-volatile information storage[3,10,11], spin transfer nano-oscillators[12,13] and logic devices[14,15].

Skyrmions were first observed in B20 materials in which the Dzyaloshinskii-Moriya interaction (DMI) originates from a non-centrosymmetric crystalline structure and leads to the formation of the intriguing spin texture[5,16-26]. In these B20 materials, the skyrmions are Bloch-type and can only exist at relatively low temperatures[4,27,28]. To create room temperature skyrmions suitable for industrial applications, hetero-structured magnetic thin-films were developed, in which the DMI is produced by breaking of the inversion symmetry at the interfaces of ferromagnetic layers and heavy metal layers with large spin–orbit coupling[27,29-37]. This interfacial DMI can result in the formation of the Néel-type skyrmions, which has been widely observed in various thin-film stacks at room temperature[27,29-37]. Furthermore, the formation and dynamics of the Néel-type skyrmions driven by the current have been investigated, which manifested an important step towards application of skyrmions in devices[9,11,33,35,38]. Usually, the out-of-plane magnetic field is very important for stabilization of skyrmions, thus, the study of the out-of-plane magnetic field on the skyrmions became one of the foci in this area[4,28]. However, the magnetization reversal and its dynamics are also governed by the in-plane magnetic field[9,36,39,40], especially for the films with very weak perpendicular magnetic anisotropy (PMA). Consequently, the study of the effects induced by the in-plane magnetic field has fundamental importance.

Lorentz transmission electron microscopy (L-TEM) is one of the most direct methods with high spatial resolution to observe the magnetic domain structures, domain walls, and skyrmions, especially for objects with the Bloch-type spin rotation. Moreover, the chirality of the Bloch-type skyrmions can also be determined by the L-TEM images[4,22,23,25,26,41], while observation of the Néel-type domain walls and skyrmions in the materials with PMA requires tilting of the sample and does not permit identification of the chirality of skyrmions from the L-TEM images alone[36,42]. In this work, we report the determination of the chirality of Néel-type skyrmions from the magnetization reversal behavior in [Pt/Co/Ta]$_{11}$ multilayers with weak PMA by introducing the in-plane magnetic field in the L-TEM imaging. We find that the in-plane magnetic field contributes to the creation of skyrmions and that high concentration of skyrmions can be achieved by increasing the in-plane field.

Our most remarkable finding is that the vector of the speed *V* with which Néel skyrmion snakes are growing in a certain direction (or the vector indicating that direction), is determined by the vector of the in-plane magnetic field $\boldsymbol{H}_{in}$, implying the relation $\boldsymbol{V} = p\, \boldsymbol{H}_{in}$ between the directions of the two vectors, with *p* being a scalar function. For most physical systems such a relation is prohibited by symmetry because *V* changes sign under spatial reflection, *r* ➔ -*r*, while $\boldsymbol{H}_{in}$ does not. One must have a pseudo-scalar in the system, that is, a scalar function *p* that changes sign under spatial reflection, in order to have the direction of the velocity determined by the direction of the magnetic field. In our system that pseudo-scalar originates from the chirality of the skyrmion that is determined by the sign of the Dzyaloshinskii-Moriya constant *D* in a non-centrosymmetric crystal. Thus, observation in which direction the skyrmion snakes are growing allows one to extract the skyrmions type and the sign of *D*.

**Characterization of the samples.** Figure 1(a) shows the high resolution, high-angle annular dark-field scanning transmission electron microscopy (HADDF-STEM) image of the cross-section of the multilayer, which exhibits a clear stack structure. Electron energy loss spectroscopy (EELS) mapping spectrum (Fig. 1(b)) further reveals a periodic tri-layered structure showing that each tri-layer is composed of a Co layer sandwiched by Pt and Ta layers. Magnetic properties of the multilayers were characterized using a SQUID-VSM magnetometer at room temperature. Figure 1(c) shows the normalized out-of-plane and in-plane hysteresis loops, which indicate that the sample possesses a weak PMA with an in-plane saturation field $\mu_0 H_k$ of only 0.206 T. This observation suggests that the sample should be very sensitive to the in-plane field. Moreover, both loops show almost zero remanence. To explore the difference in the magnetic structures in the remanent states after the sample was saturated in the presence of out-of-plane and in-plane fields, the magnetic force microscopy (MFM) measurements were performed at zero fields after each hysteresis loop measurement. The corresponding MFM images are shown in Fig. 1(d) and (e). As expected for the films with both PMA and DMI, a typical labyrinth domain structure is clearly seen after saturation by a perpendicular field. Interestingly, aligned and stripe-like domains, sharply different from the labyrinth domain structure, were observed in the remanent state after the in-plane saturation, although the magnetization within the domains was still perpendicular to the film plane.

To understand the effect of the in-plane field on the magnetization reversal in the multilayer sample, in situ L-TEM measurements were performed. In L-TEM mode, the magnetic field parallel to the electron beam can be easily applied through the objective lens of the TEM. By tilting the sample, an in-plane field can also be applied. The ratio of the in-plane and out-of-plane field can be tuned by varying the tilt angle ($\theta$)[36,42]. For the Bloch-type skyrmions, it is not necessary to tilt the sample to create the contrast in the L-TEM images. Depending on the sign of $D$, the core (shell) shows bright (dark) or dark (bright) contrast[25]. However, for the Néel-type skyrmions, the sample tilting is essential[36,42]. Here, we introduce briefly the physics of the L-TEM imaging for Néel-type skyrmions. Figure 1(f) shows the top view schematic diagram of the L-TEM imaging. The sample is tilted from the *xy* plane to the *x'y'* plane. Blue, white and red contrasts represent the outside, the boundary and the inside of a skyrmion with negative, zero and positive magnetization along the *z'* axis. The corresponding *xy* plane projection of the magnetization is indicated with the brown arrows. The green arrows indicate the corresponding Lorenz force when electrons pass through the skyrmion. It is clear that the skyrmion edge does not show any intensity contrast in the L-TEM image because the intensity change caused by the Lorenz force at any point in this zone is compensated or cancelled. The intensity contrast is formed in the L-TEM images due to the contributions from both the outside and the core of the skyrmion. Therefore, the resulted image for the skyrmion should be of dark-bright contrast with dark at the top side and bright at the bottom side just under the Néel-skyrmion edge, as is shown in the right side of Fig. 1(f). The distance between the intensity minimum and the density maximum represents the skyrmion size.

**Magnetization reversal under a tilted magnetic field.** We start with the tilting angles of $\alpha = 23°$ and $\beta = -20°$ for the L-TEM imaging. Here, $\alpha$ is the tilting angle around the *y* axis and $\beta$ around the *x* axis. The *z* axis is chosen to be parallel to the TEM cylinder. Thus, the total tilt angle $\theta = 30°$ ( $\cos(\theta) = \cos(\alpha) \cdot \cos(\beta)$ ). Before imaging, the sample was saturated in the down direction (-*z*) while the direction of the in-plane (*x'y'* plane) component of the magnetic field pointed to the top left as indicated by the arrow in up-right corner of Fig. 2(a). The evolution of the domain structure was imaged at the defocus of 7.62 mm by varying the magnetic field from the negative saturation field to the positive saturation field as is shown in Fig. 2 (a) (See the detailed reversal process in Supplementary Movie 1). When the magnetic field was changed to about -1648 Oe (the out-of-plane component $H_{out}$=1427 Oe and in-plane component $H_{in}$= 824 Oe), an isolated skyrmion emerged as marked by the green circle with the contrast of bright at upper right and dark at lower left. By continually varying the field, more skyrmions appeared (such as at -1600 Oe) and began to evolve into snake-like structures as is shown in the image obtained at -1464 Oe. The maximum skyrmion density before the formation of the snake-like structures is denoted as $d_{max1}$. By changing the field further (from -1464 Oe to -1320 Oe), the length of the snake-like structures increases. Meanwhile, some new skyrmions appeared (as indicated by yellow arrows in the image obtained at -1456 Oe) and rapidly extended to the snake-like structures. We compared the two images obtained at -1468 Oe and -1456 Oe and marked the structure differences with green dashed ellipses. Interestingly, we found that the snake-like structures mainly extended along the directions that are opposite to the direction of the in-plane field. Finally, the density of the snake-like structures achieved maximum over the entire film at about -680 Oe, forming an almost aligned arrangement. When the field approached zero, the in-plane field became zero too, the oriented stripe domains became less aligned. With increasing the field in the positive direction the in-plane field increases in the opposite direction. The stripe domains become better oriented again as seen in the image obtained at 1080 Oe. When increasing the field to 1400 Oe, the orientated domain structures fractured and a mixed state of skyrmions and snake-like structures appeared. Eventually, all snake-like domains broke into skyrmions at 1720 Oe. The maximum skyrmion density obtained in this image was defined as $d_{max2}$. With increasing the field further, the skyrmions annihilated gradually (1840 Oe) and the film was finally saturated to a ferromagnetic state at 1960 Oe.

**In-plane field induced anisotropic growth of the snake-like structures.** To explore the correlation between the in-plane field direction and the preferred extension direction of the snake-like domains, more experiments were performed with different tilting angles and directions, and same phenomena were observed. Note that the extension of the snake-like structures is driven by the out-of-plane component of the field, but their anisotropic formation is induced by the in-plane field. Figures 2(b) and 2(c) show L-TEM images of the extending snake-like structures taken during changing the magnetic field from negative saturation field to zero with the tilting angles of $α = 0°; β = -20°$ and $α = -28°; β = -20°$ respectively. The corresponding directions of in-plane fields were indicated by black arrows. Similarly, we indicated the changes in the images with green dashed ellipses. It is clear that for both Fig. 2(b) and 2(c), the snake-like structures preferred to grow along the directions opposite with that of the in-plane fields and finally formed an almost aligned arrangement.

**Relationship between in-plane field direction and the preferred extension direction.** The observation of the preferred extension direction governed by the in-plane field may become a useful and reliable approach to determine the skyrmion chirality, which is of great importance for manipulating skyrmions as well as for establishing the sign of the DMI constant[4]. To further understand this phenomenon and also determine the chirality of skyrmions, micromagnetic simulations were performed with the mumax$^3$ software[43]. As an example, we first studied a system with an interfacial DMI of positive $D$. Following the experimental protocol, we created some skyrmions randomly and stabilized them with a magnetic field applied in the -$z$ direction. Figure 3(a) shows the evolution of the magnetic structure when decreasing the magnetic field with the tilt angles of $θ = 0°$ and $θ > 0°$. In the case of $θ = 0°$, when no in-plane field is applied, the skyrmions extended isotropically, leading to a typical labyrinth domain structure, as we observed experimentally (Fig. 1(d)). However, at $θ > 0°$, when a non-zero in-plane field component was applied along the +$x$ axis, an anisotropic growth of the skyrmions was clearly observed. An important finding is that the snake-like structures grow parallel to the in-plane magnetic field. We also found that the growing speed in the -$x$ direction is much greater than in the +$x$ direction. Thus, we concluded that the preferred extension direction is opposite to the direction of the in-plane magnetic field, when the magnetization of the skyrmion core points up for an interfacial DMI with positive $D$.

More simulations were performed for both Bloch-type skyrmions and Néel-type skyrmions with different sign of $D$. The results are summarized in Fig. 3(b-f). Fig. 3(b) shows the spin texture of the four types of skyrmions whose magnetization at the skyrmion core is pointing up. For a Bloch-type skyrmion, the in-plane spin components rotate clockwise for positive $D$ and counter-clockwise for negative $D$, while for the Néel-type skyrmions, the in-plane component spins point inside (positive $D$) or outside (negative $D$). Based on the principle of the contrast formation in the L-TEM imaging, it is not necessary to tilt the sample to create the contrast in the L-TEM images for the Bloch-type skyrmions and their chirality could be distinguished from the contrast (bright or dark) under the skyrmion core (Fig. 3(c)), while for the Néel-type skyrmions, the sample tilting is essential for the L-TEM imaging and same contrasts (Fig. 3(c)) will be observed regardless of the sign of $D$. Therefore, their chirality cannot be identified by the normal L-TEM imaging. However, we could solve this problem by applying an in-plane magnetic field. The in-plane field (+$x$ direction here) breaks the symmetry of the magnetization distribution of a skyrmion as is shown in Fig. 3(d), because the region with spins parallel to the in-plane component of the field grows while the region with the opposite spins shrinks. Though the in-plane component of the fields is in the +$x$ direction for all the four cases, it is found that the preferred extension directions are all different as marked in Fig. 3(e). For samples with bulk DMI, the preferred extension directions are in the +$y$ or -$y$ direction depending on the sign of $D$. The final domain structure at zero field is thus an orientated structure in the $y$ directions as is shown in Fig. 3(f). This behavior has been observed experimentally[44]. For the interfacial DMI, the preferred extension direction is collinear with the in-plane field. By comparing theoretical prediction with experimental data on the relationship between the preferred extension direction and the in-plane field direction, we were able to identify the chirality of the skyrmions (or the sign of $D$). It is evident that $D > 0$ in our [Pt/Co/Ta]$_{11}$ multilayers.

**Skyrmion creation assisted by the in-plane field.** In Fig. 2, we can see that the achieved skyrmion densities are different for different tilting angles (or in-plane field). To understand this phenomenon, we increased the tilt angle $\theta$ from 3º to 45º, and studied the magnetization reversal using the in-situ L-TEM. As has been discussed in Fig. 2(a), the skyrmions appeared twice when the magnetic field was swept from negative saturation field to positive saturation filed. The first emergence of skyrmions corresponded to the nucleation from the ferromagnetic state at negative field and the second emergence corresponded to the creation of skyrmions by breaking the labyrinth domains by the positive field. Figure 4 shows the maximum densities $d_{max1}$ and $d_{max2}$ as functions of the tilt angle and it also shows the corresponding L-TEM images. One can see that when the tilt angle is very small (i.e., 3º, consequently, the in-plane field is very small), $d_{max1}$ is almost zero and the snake-like structures are created directly. However, many more skyrmions were created at the breaking process, when the field was increased from zero to positive saturation. The reason why $d_{max2}$ is much larger than $d_{max1}$ over the whole range of tilt angles is that the breaking process is energetically more favorable than the nucleation process. Interestingly, with increasing the tilt angle (increasing the in-plane field), we found that both $d_{max1}$ and $d_{max2}$ increase, which indicates that the in-plane field favors the creation of skyrmions. This can be understood in the following way. It has been shown that skyrmion density increased with increasing the critical material parameter $\kappa = \pi D / 4\sqrt{AK_{eff}} = D/D_c$ [34,37], where $A$ is the exchange stiffness, $K_{eff}$ is the effective perpendicular anisotropy and $D$ is the DMI constant. Application of the in-plane field will actually weaken the role of the perpendicular anisotropy, while keeping $D$ and $A$ unchanged, which leads to the increase of $\kappa$.

In summary, we deposited [Pt/Co/Ta]$_{11}$ multilayer sample with weak PMA and investigated the magnetization reversal behavior for different tilted angles of the field, to have both magnetic field components parallel and perpendicular to the film plane, using in-situ L-TEM. We found that the chirality of skyrmions can be identified by their preferred extension direction under an in-plane magnetic field, which was confirmed by micromagnetic simulations. Furthermore, we found that the in-plane field facilitates the creation of skyrmions and increases their concentration.

## Methods:

**Sample preparation:** The PMA multilayer stacks Ta(5 nm)/[Pt(4 nm)/Co(1.4 nm)/Ta(1.8 nm)]$_{11}$ were deposited simultaneously both on thermally oxidized Si substrates and carbon membranes by DC magnetron sputtering at room temperature. The pressure of the Ar gas was 0.4 Pa. The base pressure was lower than ~2×10$^{-5}$ Pa. The films grown on thermally oxidized Si substrates were used for the SQUID-VSM and MFM measurements. The films on carbon membranes were used for the in-situ L-TEM measurements.

**Transmission electron microscopy (TEM):** TEM sample was prepared using focused ion beam (FIB) from the sample deposited on thermally oxidized Si substrates. The High angle annular dark field scanning TEM (HADDF-STEM) images and Electron energy loss spectroscopy (EELS) mapping spectrum were employed to analyze the structure of sample in a FEI TEM (Titan 60-300).

**L-TEM measurements:** In-situ Lorentz transmission electron microscopy (L-TEM) imaging was carried out by using a FEI Titan Cs Image TEM in Lorentz mode (the Fresnel imaging mode) at 300 kV. A double holder was used for the measurements. The magnetic field parallel to the electron beam can be easily applied and tuned by varying the current passing though the microscope's objective lens. By tilting the sample, an in-plane field can also be applied. The ratio of the in-plane and out-of-plane field can be tuned by varying the tilt angle ($\theta$).

**MFM measurements:** The magnetic force microscopy (MFM) experiments were performed using an Agilent 5500 Scanning Probe Microscope in tapping/lift mode.

**Micromagnetic Simulations:** mumax$^3$ software package, including the extension module of the DMI, was used for the micromagnetic simulations. A 1×1 μm$^2$ square system with 2D periodic boundary conditions was used as a material system. The mesh size is set to 2×2 nm$^2$ and the following parameters were chosen: $M_s$ = 7.0×10$^5$ A/m, $K_u$ = 3.5×10$^5$ J/m$^3$, $A$ = 1.0×10$^{-11}$ J/m and $T$ = 0 K. For comparison, both bulk DMI and interfacial DMI were considered and its magnitude was set to $D$ = 1.3×10$^{-3}$ J/m$^2$.

# ACKNOWLEDGEMENTS

This publication is based on research supported by the King Abdullah University of Science and Technology (KAUST), Office of Sponsored Research (OSR) and under the award No. OSR-2016-CRG5-2977.

Figure 1

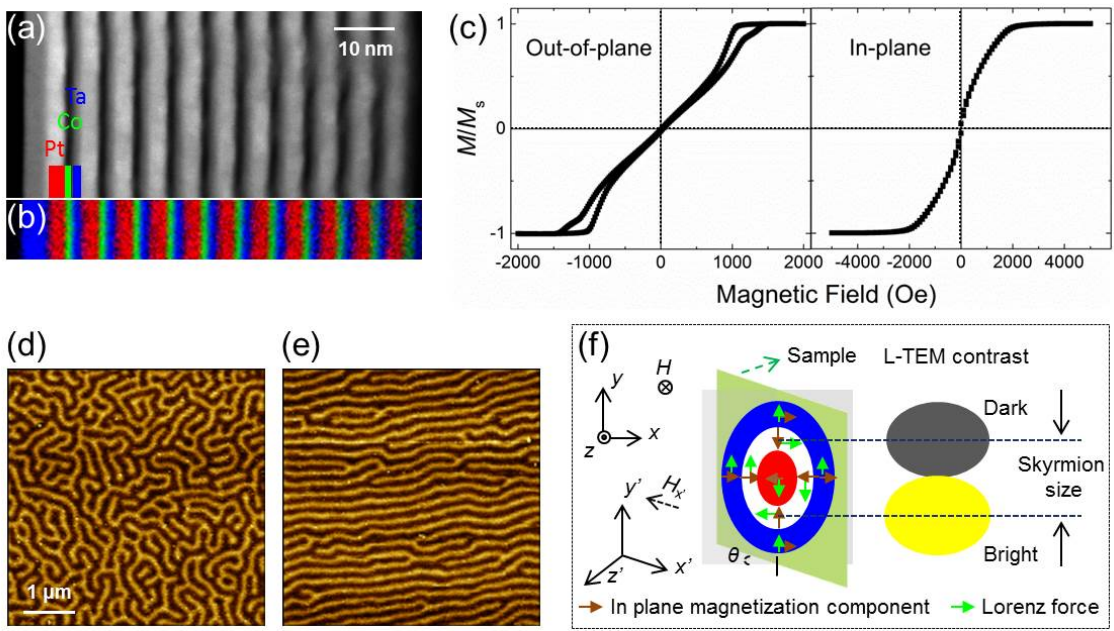

Figure 1 **Structure analysis and magnetic properties.** (a) HADDF-STEM image and (b) EELS mapping spectrum of the cross-section of the Ta(5 nm)/[Pt(4 nm)/Co(1.4 nm)/Ta(1.8 nm)]$_{11}$ multilayer sample. (c) Normalized out-of-plane and in-plane hysteresis loops of the sample. MFM images at zero fields after the (d) out-of-plane and (e) in-plane saturation fields. (f) Schematic diagram of a Néel-skyrmion on a tilting sample for L-TEM imaging. The blue/red contrast represents areas with negative/positive magnetization along *z'*. The corresponding *xy* plane projection is indicated with the brown arrows. The green arrows indicate the Lorenz force when electrons pass through the skyrmion. The expected Fresnel contrast is shown at the right side and the distance between the intensity minimum and density maximum represents the skyrmion size.

Figure 2

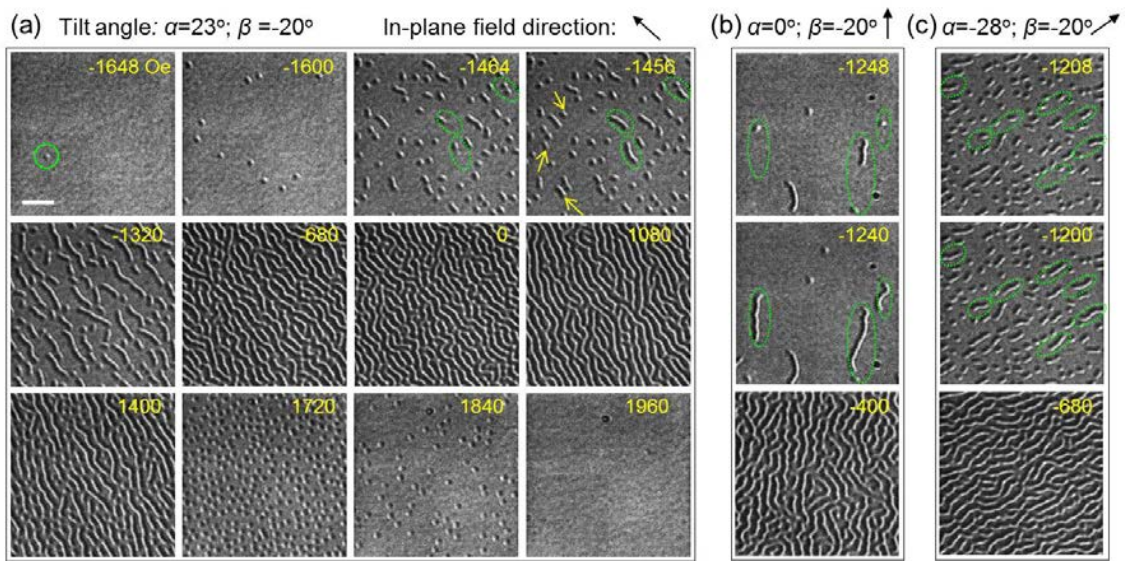

**Figure 2. In-situ L-TEM observation of reversal behaviors.** (a) In-situ L-TEM observation of the [Pt/Co/Ta]$_{11}$ multilayer's reversal behaviors. The images were taken at a tilt angle of $\alpha = 23^o$ and $\beta = -20^o$ and the direction of in-plane component field is marked by black arrow. L-TEM images of the snake-like structures extension taken during changing the magnetic field from negative saturation field to zero with the tilting angles of (b) $\alpha = 0^o$; $\beta = -20^o$ and (c) $\alpha = -28^o$; $\beta = -20^o$. The scale bar corresponds to 1 μm.

Figure 3

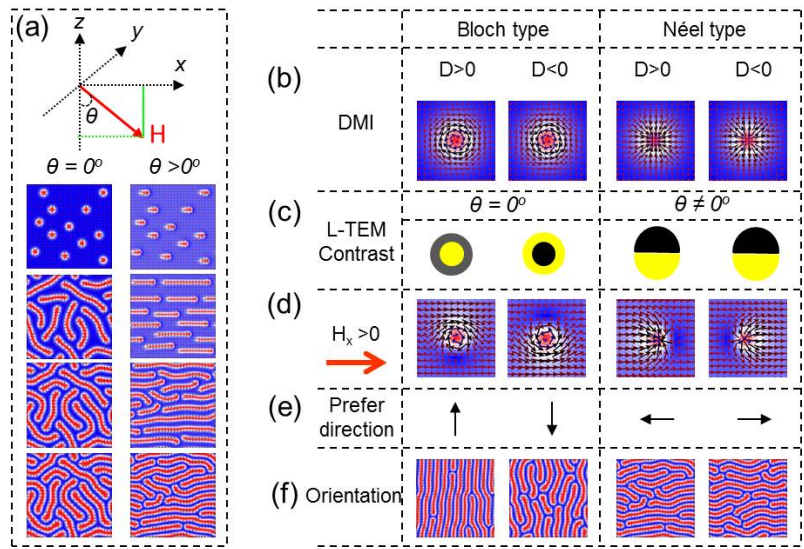

**Figure 3. Micromagnetic simulations.** (a) Magnetic structure evolution when increasing magnetic field from negative saturation to zero with a tilt angle of $\theta = 0$ and $\theta > 0$. (b) Four kinds of skyrmions and (c) the expected L-TEM contrast for Bloch-type skyrmion without tilting the sample and for Néel-type skyrmion on a tilt sample. (d) Magnetic structure when apply an in-plane field. (e) The preferred direction when the in-plane field is in the +x direction. (f) The final orientated domain structure at zero fields.

Figure 4

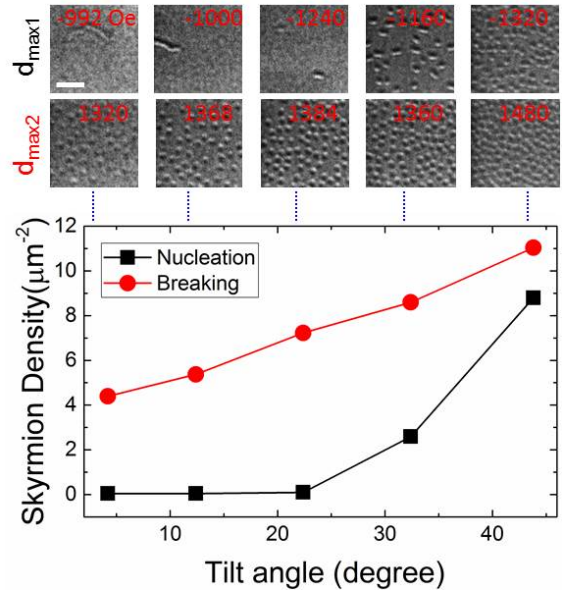

**Figure 4. Relationship between skyrmion densities and tilt angles.** The maximum skyrmion densities $d_{max1}$ and $d_{max2}$ as functions of the tilt angles. The top shows the corresponding L-TEM images. The scale bar corresponds to 1μm.